\documentclass[aps,pra,reprint,floatfix,superscriptaddress]{revtex4-1}
\usepackage{amsmath}
\usepackage{graphicx}
\usepackage{booktabs}
\usepackage{textcomp}
\usepackage{dcolumn}
\usepackage{sidecap}
\usepackage{subfigure}

\begin{document}
\title{Capture and isolation of highly-charged ions in a unitary Penning trap}

\author{Samuel M Brewer}
\altaffiliation[Current Address: ]{Time and Frequency Division, NIST, Boulder CO, 80305}
\affiliation{University of Maryland, College Park, Maryland 20742, USA}
\author{Nicholas D Guise}
\affiliation{University of Maryland, College Park, Maryland 20742, USA}
\affiliation{National Institute of Standards and Technology, 100 Bureau Drive,
Gaithersburg, Maryland 20899-8422, USA}
\author{Joseph N Tan}
\affiliation{National Institute of Standards and Technology, 100 Bureau Drive,
Gaithersburg, Maryland 20899-8422, USA}

\date{\today}

\begin{abstract}
We recently used a compact Penning trap to capture and isolate highly-charged ions extracted from an electron beam ion trap (EBIT) at the National Institute of Standards and Technology (NIST). Isolated charge states of highly-stripped argon and neon ions with total charge $Q \geq 10$, extracted at energies of up to $4\times 10^3\,Q$ eV, are captured in a trap with well depths of $\,\approx (4\, {\rm to}\, 12)\,Q$ eV.  Here we discuss in detail the process to optimize velocity-tuning, capture, and storage of highly-charged ions in a unitary Penning trap designed to provide easy radial access for atomic or laser beams in charge exchange or spectroscopic experiments, such as those of interest for proposed studies of one-electron ions in Rydberg states or optical transitions of metastable states in multiply-charged ions.  Under near-optimal conditions, ions captured and isolated in such rare-earth Penning traps can be characterized by an initial energy distribution that is $\approx$ 60 times narrower than typically found in an EBIT.  This reduction in thermal energy is obtained passively, without the application of any active cooling scheme in the ion-capture trap.
\end{abstract}

\maketitle

\section{Introduction}
\label{sec: intro}
Highly-charged ions (HCI) are of interest in the study of atomic structure, astrophysics, and plasma diagnostics for fusion science \cite{HCI}.  The high nuclear charge, $Z$, tends to amplify relativistic effects in atoms, such as fine and hyperfine structure splitting \cite{JDGrev2001}.  For example, the fine structure energy splitting is proportional to $(Z \alpha)^4$, where $\alpha \approx 1/137$ is the fine structure constant, and hence can be so large for some high $Z$ ions that the transition frequency is scaled up from the microwave to the visible domain of the electromagnetic spectrum \cite{Mohr2008} -- a useful feature for observing astrophysical objects.

Apart from natural sources, highly-charged ions have become more widely accessible with the development of laboratory facilities like heavy-ion storage rings \cite{mpq-tsr} and more compact devices like the electron-cyclotron resonance (ECR) ion source \cite{ecris} and the electron beam ion trap/source (EBIT/EBIS) \cite{ebit1,Donets1998rsi,motohashi:890,xiao:013303}.  These ion sources are useful in various research areas, including: spectroscopy (moments, spectral lines, etc.), ion-surface interactions \cite{HCISP}, plasma diagnostics for next-generation tokamak fusion reactors such as the International Thermonuclear Experimental Reactor (ITER) \cite{pra09}, and tests of astrophysical models (see \cite{fe17ebit} and references therein).  

The isolation of single species, highly-charged ions at low energy in traps can enable some interesting studies of atomic and nuclear phenomena \cite{Church1993}.  As a recent example, high precision studies of HCIs have been proposed to realize atomic clocks based on Nd$^{13+}$ and Sm$^{15+}$ \cite{HCIalphadot} for laboratory investigations of the variation (temporal and spatial) of $\alpha$.  Another possibility is to test theory in Rydberg states of one-electron ions with comb-based spectroscopy, which could led to a Rydberg constant determination that is independent of the proton radius. \cite{Mohr2008}

A broad survey of trap types and ion sources developed to advance measurements of atomic and nuclear properties can be found in the 2003 review article by Kluge, {\it et al.} \cite{Kluge2003}.  A variety of useful techniques have been developed for the study of trapped positrons \cite{Surko2004}, antiprotons \cite{atrapiont} and antihydrogen (see Ref. \cite{hbar2012} and references therein) as well as highly-charged ions in Penning traps \cite{Penn} with meter-long electrode structures surrounded by multi-tesla solenoid magnets \cite{spectrap2010, pentatrap2012}.  In some of the earliest experiments, a cryogenic Penning trap (RETRAP) with a high-field superconductive magnet \cite{retrap1} was employed to capture ions extracted from an EBIT at the Lawrence Livermore National Laboratory (LLNL).  More recently, SMILETRAP II demonstrated capture and cooling of Ar$^{16+}$ in a Penning trap utilizing a room-temperature 1.1 T solenoid magnet \cite{smilePRL2011}.  

Solenoidal magnets can generate a strong magnetic field for ion confinement, but they also impose geometrical constraints that hinder the access of laser or atomic beams to be directed at the stored ions.  In our effort to produce and study one-electron ions in Rydberg states, we have designed unitary Penning traps for isolating single-species charge states of highly-stripped ions extracted from an EBIT at NIST \cite{unitpen}.  The unitary architecture is useful also for studying long-lived transitions, as will be discussed in forthcoming publications.  Initial demonstrations \cite{unitpen, HCI2010, CDAMOP2011} reported the use of unitary Penning traps to isolate and store various HCIs.  In this work, we discuss the dynamical considerations and experimental manipulations that are essential for optimized performance to maximize the number of stored ions as well as minimize the energy distribution for precise measurements.

A brief description of the system configuration is provided in Sec.\,\S \ref{sec: expsu}.  Numerical simulations were carried out to guide the design of the compact Penning trap and additional beam-conditioning components, as discussed in Sec.\,\S \ref{sec: ioncapsim}, with emphasis on the deceleration of fast ($\approx$ 40 keV) ions approaching the region $\approx$ 3 cm in front of the trap.  Section\,\S \ref{sec: EBIT} describes charge state selection and ion pulse optimization, emphasizing the importance of (a) minimizing the time width of the extracted ion pulse, and (b) matching the deceleration potential near the Penning trap to ion extraction energy.  Results from recent ion capture experiments are presented, illustrating ion capture optimization (Sec.\,\S \ref{sec: ioncap}) and residual energy measurement (Sec.\,\S \ref{sec: ionceng}). Finally, a discussion of the ion capture efficiency is presented in Sec.\,\S \ref{sec: capeff}.

\section{Experimental Setup}
\label{sec: expsu}

\begin{figure*}
\begin{center}
\includegraphics[angle=90,width=1.0\textwidth]{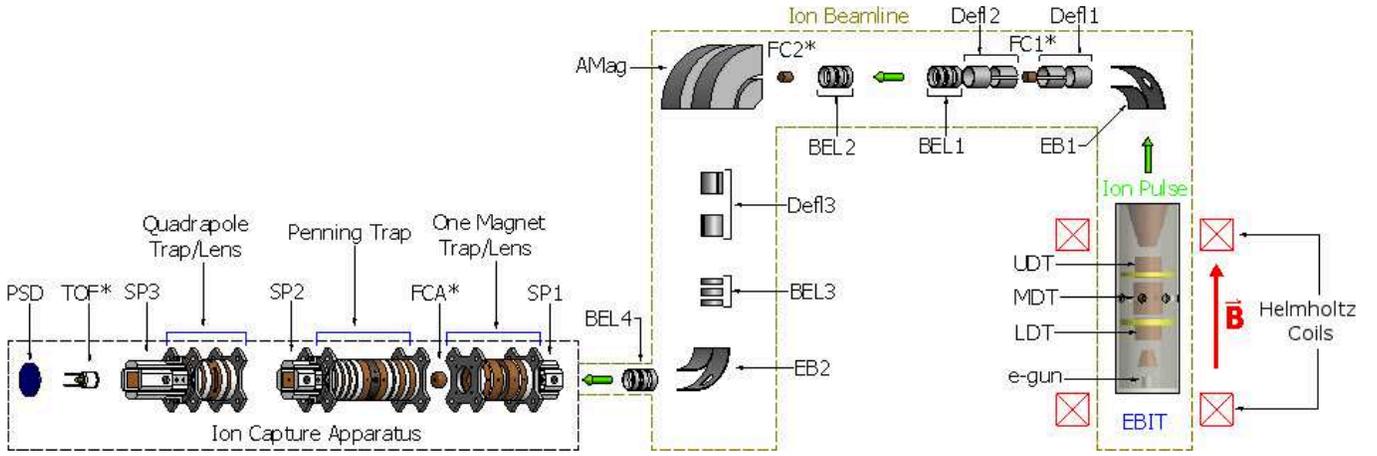}
\caption{\label{fig: ebitapp}(Color online) Schematic overview of the experimental set-up (NOTE: Not to scale).  The ion source is the EBIT at NIST with its existing ion extraction beamline, which has an analyzing magnet (AMag) for charge state selection.  The experiment apparatus at the end of the beamline houses a unitary Penning trap to capture selected ions, and detectors to count ions ejected after storage.  Labels with asterisk indicate mounting on retractable translators.   Broken lines represent the boundary of evacuated space; vacuum pumps are not shown.  Ion-trajectory path-length from the EBIT to the Penning trap is $\approx$ 8 meters.}
\end{center}
\end{figure*}

The experimental set-up, illustrated in Figure \ref{fig: ebitapp}, consists of the EBIT with its ion extraction beamline, and the recently-installed ion-capture apparatus.
 Since some parts of the set-up have been described in detail elsewhere \cite{unitpen,CDAMOP2011}, only a brief overview is given here.

Highly-charged ions are produced in the EBIT, bound radially to the energetic electron beam along the axis.  Axially the ions are trapped in an electrostatic well created by applying electric potentials $(V_i)$ to three cylindrical electrodes, called drift tubes--labelled by their location: upper (UDT), middle (MDT), and lower (LDT), with $V_{\rm MDT} < V_{\rm UDT} < V_{\rm LDT}$.  To extract an ion bunch, the MDT can be quickly raised to a value $V_{\rm UDT} < V_{\rm MDT} < V_{\rm LDT}$ thus ejecting HCIs into the beamline \cite{HCI2010,PikinBeamline,RatliffEBITBeamline}.

Electrostatic ion optics in the beamline guide (EB1, Defl 1-3, EB2) and focus (BEL 1-4) the extracted ions, transporting them over an 8-meter trajectory from the EBIT to the unitary Penning trap.  At various points, retractable Faraday cups (FC1-2) can be  inserted to monitor the ion beam.  About half-way along the beamline, an analyzing electromagnet (AMag) selects a specific charge state to be captured in the Penning trap.  The beamline vacuum space has a base pressure of $2.7 \times 10^{-7}$ Pa ($2.0 \times 10^{-9}$ Torr).  

\begin{figure}
\begin{center}
\includegraphics[angle=90,width=8.6cm]{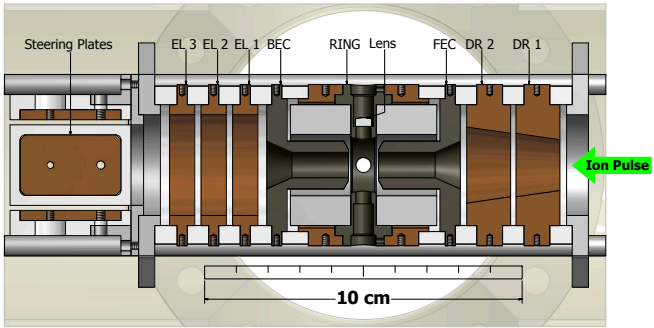}
\caption{\label{fig: trapapp}(Color online) Cross-sectional view of the compact Penning trap used to capture ions (foreground). The ring electrode has 4 equidistant holes--one hole concentric with a vacuum window in the background; a small lens is inside the top hole for observing fluorescence from stored ions. Two rare-earth (NdFeB) magnets are embedded within the electrode assembly, one on each side of the ring electrode.  Ions enter from the right-hand-side along the trap axis, slowing in the deceleration electrodes (DR1 and DR2) before entering the trap via the 8.00 mm hole in FEC.  Stored ions can be counted by ejection to a TOF detector, focussed and guided by an einzel lens (EL 1, EL 2, EL 3) and steering plates.}  
\end{center}
\end{figure}

At the entrance of the ion-capture apparatus, specialized components are used to optimize on-axis injection of HCIs into the unitary Penning trap; a set of four steering plates (SP1), a one-magnet trap/einzel lens, and a retractable Faraday cup (FCA) allow fine adjustments in alignment and ion pulse conditioning \cite{CDAMOP2011}.  After confinement in the Penning trap (Fig. 2), stored ions are detected by ejection to one of the ion detectors. A retractable micro-channel plate (MCP) with fast response is used for ion counting and time-of-flight (TOF) or charge state analysis.  If the fast TOF detector is retracted, as discussed in \cite{CDAMOP2011}, a position-sensitive MCP ion detector (PSD) can be used during beam alignment and conditioning.  The TOF detector is a ``Chevron", or V-stack type \cite{RSIMCP1973} with a disc head (8.0 mm active diameter), which is operated in either proportional (charge amplifying) mode, or in a fully-saturated, event counting mode.  An event pulse has rise/fall time $\approx$ 350 ps with a gain of $> 10^6$ per incident charge.  

Figure \ref{fig: trapapp} shows a half-cut view of the unitary Penning trap used to capture ions. The unitary architecture \cite{unitpen} makes the ion trap extremely compact, with an electrode assembly volume of less than 150 cm$^3$. The magnetic field for radial confinement of stored ions is generated by two rare-earth magnets that are yoked by the soft-iron electrodes (FEC, RING, and BEC).  The front endcap (FEC) and the back endcap (BEC) are maintained at a higher potential than the RING electrode to form an axial trapping well. The two deceleration electrodes (DR 1 and DR 2) adjacent to the front endcap are crucial for slowing ions before they enter the unitary Penning trap; their conical inner surfaces are tailored to produce near-planar equipotential surfaces.  Application of static and time-varying electrical potentials is controlled through a computer interface, the details of which are provided in Sec. \ref{sec: expres}.  A separate vacuum chamber houses the room-temperature Penning trap, allowing control of the background gas composition and pressure; the base pressure of this vacuum chamber is $1.0 \times 10^{-7}$ Pa ($7.6\times 10^{-10}$ Torr).  

\section{Simulations}
\label{sec: ioncapsim}

Numerical simulations have been carried out to investigate: (a) the optimal electrode geometry of a unitary Penning trap designed to slow, capture, and store ions extracted from an EBIT; (b) the operation settings, such as voltages and switching times for controlling electrodes; and (c) the ideal conditions of an incoming ion bunch. Ion capture simulations involve computations of both the magnetic field in the trap as well as the electrostatic potential generated by the trap electrodes and focusing elements, generally under time-varying potentials.  The details of the magnetic field calculations, including comparisons with measured trap fields, are presented in \cite{unitpen}.  The measured magnetic field strength is $\approx$ 310 mT in the trapping region and is in good agreement with the calculated field.  The electric field in the trap assembly is calculated using a numerical Boundary Element Method (BEM), originally developed for computing properties of electrostatic lenses \cite{FRead}. 

\begin{figure}
\begin{center}
\includegraphics[angle=0,width=0.4\textwidth]{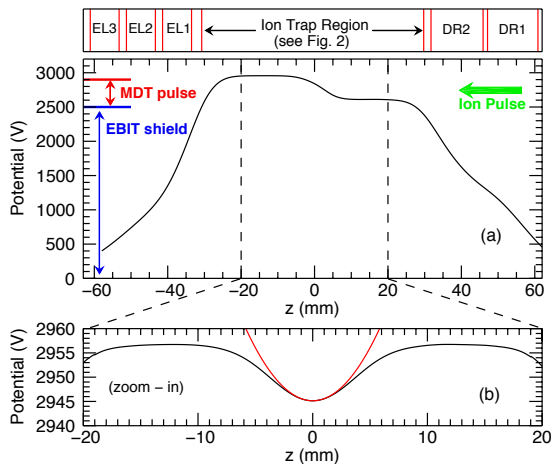}
\caption{\label{fig: vaxoc}(Color online) Calculated electrostatic potential along the trap axis with the electrode positions indicated at the top of the figure. The ``open" trap condition is shown in part (a), with the EBIT shield voltage and the MDT pulse voltage indicated. The ``closed" trap condition is shown in part (b), magnified near the trap center at z = 0 mm, with BEM calculation in black, and analytic quadrupole fit in red.   The ion pulse enters the apparatus from the right. The applied voltages are given in Table \ref{tab: param}; the difference of 30 V between FEC $=$ BEC and the ring electrode corresponds to an on-axis well depth of $11.64$ V.}
\end{center}
\end{figure}

An example of the calculated electrostatic potential along the axis of the ion trap is shown in Fig. \ref{fig: vaxoc}.  The ``open" condition in preparation for ion capture is shown in (a) and the ``closed" condition following ion capture is shown in (b).  The applied voltages for each electrode and the critical EBIT parameters are listed in Table I.  The EBIT shield voltage and MDT high voltage pulse levels are included in Fig. \ref{fig: vaxoc}a for comparison.  As shown in Fig. \ref{fig: vaxoc}b, the axial potential well near the trap center is well approximated by an analytic quadrupole potential, which in cylindrical coordinates takes the form \cite{gabrev,unitpen}

\begin{equation}
\label{pennV}
V = \lambda V_0 \frac{z^2 - \rho^2/2}{2d^2}  + V_C  .
\end{equation}
The field coordinates $z$ and $\rho$ are defined from the center of the trap; $V_0$ is the applied potential difference between the endcaps and the central ring electrode, $V_C$ is the common-mode or float potential, and $d$ is a geometric factor 
\begin{equation}
\label{d2}
d^2 \equiv \frac{1}{2} (z_0^2 + \rho_0^2/2)  .
\end{equation}
The coefficient $\lambda$ (often referred to as $C_2$) is of order unity.  The characteristic dimensions $r_0$ and $z_0$ are from the center of the trap to the ring and endcap electrodes, respectively.  For the Penning trap presented here, $\rho_o = 8.5$ mm, $z_o = 4.736$ mm, and $\lambda = 0.854$.

\begin{table}[htdp]
\begin{center}
\begin{tabular}{|c|c|}
\hline
\multicolumn{2}{|c|}{Penning Trap Parameters} \\
\hline
Trap Electrode & Applied Potential (V) \\
\hline
DR1 & 1300.0 \\
DR2 & 1600.0 \\
FEC & (Low) 2610.0 \\
         & (High) 2956.8 \\
Ring & 2926.8 \\
BEC & (Low) 2460.0  \\
         & (High) 2956.8 \\
EL1 & 500.0 \\
EL2 & 1500.0 \\
EL3 & 500.0 \\
\hline \hline
\multicolumn{2}{|c|}{EBIT Parameters} \\
\hline
e$^{-}$ beam Energy & 2.5 keV \\
e$^{-}$ beam Current & 14.4 mA \\
LDT & 500 V \\
MDT & Trap Dump = 400 V \\
UDT & 220 V \\
Ionization Time & 76.0 ms \\
Analyzing B-field & 66.22 mT \\
\hline
\end{tabular}
\end{center}
\caption{\label{tab: param}Typical applied trap potentials and EBIT parameters used in producing and capturing Ar$^{13+}$ ions.  The EBIT conditions have been chosen to both maximize ion production and minimize the time width of ion pulse.}
\vspace{0.07cm}
\end{table}

Special care was taken in designing the two deceleration electrodes, DR1 and DR2, to generate nearly planar equipotential surfaces with resulting $\nabla \Phi$ gradient that tends to remove axial kinetic energy from ions entering the trap.  In order to attain the lowest possible residual energy after capture it is important to minimize momentum transfer to transverse motions as the ions are injected into the Penning trap.  

With the computed electric and magnetic fields \cite{unitpen} and a given set of initial conditions (the ion position and velocity), an ion trajectory is calculated by integrating the equations of motion using an adaptive step-size Runge-Kutta technique such as provided by a commercial code, Charged Particle Optics \cite{FRead,disclaimer}.  A triangle mesh ratio limit (side/length) of 20 yields fractional precision of 10$^{-4}$ for the electric field and ray tracing computations.

\begin{figure}
\begin{center}
\includegraphics[angle=0,width=0.4\textwidth]{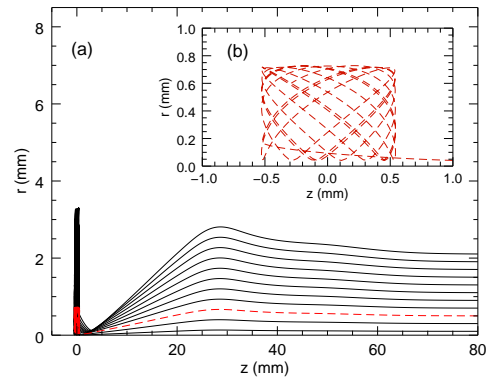}
\caption{\label{fig: simtraj}(Color online) Classical single-ion trajectories computed for a family of impact parameter values, a$_i$, ranging from 0.1 mm to 2.1 mm in 0.2 mm steps.  The ion trap center is located at z = 0 mm.  For the same velocity parallel to the axis, the amplitudes of bound ion motions increase with increasing a$_i$.  The trajectory shown in red dotted line, magnified in the inset (b),  corresponds to an impact parameter of 0.5 mm.}
\end{center}
\end{figure}

\begin{figure}
\begin{center}
\includegraphics[angle=0,width=0.4\textwidth]{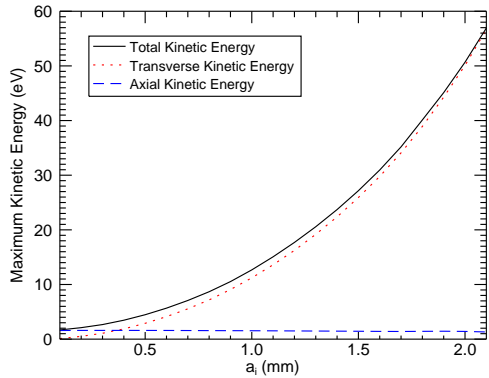}
\caption{\label{fig: simeng1}(Color online) Maximum kinetic energy of captured Ar$^{13+}$ ions, calculated as a function of the impact parameter, a$_i$.  Ions enter the capture apparatus with velocity parallel to the axis.  The total kinetic energy is shown as a solid line (--), the transverse kinetic energy is shown as a dotted line ($\cdots$), and the axial kinetic energy is shown as a dashed line ($--$).}
\end{center}
\end{figure}

In this work, only single particle trajectories are computed to model the properties of the system.  An improved model would  require the inclusion of the inter-ion coulomb interaction, and is not practical for computational resources available in this work.  To first approximation, single-particle trajectories have been useful in finding the optimal conditions for successful ion capture. To illustrate, trajectories calculated for a range of impact parameter values, $a_{i}$ (perpendicular distance from trap axis at $z > 70$ mm) are presented in Fig. \ref{fig: simtraj}.  Each trajectory starts with the same initial velocity entirely parallel to the trap axis (the direction of propagation), representing the zero-emittance \cite{cpoptics} beam condition.  Iterating such computation for various trap parameters, the potentials on the deceleration electrodes DR1 and DR2, as well as the electrode geometry, have been optimized to capture ions in trajectories with the smallest amplitudes of resulting bound motions.  Fig. \ref{fig: simeng1} shows the maximum ion kinetic energy after capture, calculated as a function of impact parameter, for Ar$^{13+}$ ions ($Q\,=\,13$; Ar\,{\scriptsize XIV} in spectroscopic notation).  The deceleration is most effective on-axis, for which the initial ion kinetic energy is removed more completely.  As the impact parameter increases, the residual energy after capture increases.  

Single particle simulation has been particularly useful for finding the capture time ($t_{capture})$ at which the Penning trap must be switched from the open configuration to the closed configuration to capture and store ions. A rough estimate is the mean transit time of the ion pulse from the EBIT to the Penning trap.  The front endcap (FEC), momentarily held below the ring potential to admit ions into the trap, must be switched to close the trap within a certain arrival time tolerance. If FEC is switched to close the trap too early, before the extracted ions enter the trapping region, the ions will scatter off and not be captured. On the other hand, if FEC is closed too late, ions will have entered the trap, turned around, and exited the trapping region--before they can be captured.  For a given initial energy and trap well configuration, there is a range of arrival times wherein the FEC electrode can be switched to successfully confine the ions that have entered the trap; the width of this allowed range for ion capture is labeled the capture time width (CTW).  The CTW can be estimated by computing ion trajectories to find bound motions for a family of times at which FEC is switched to close the trap, in 10 ns time steps, assuming the same initial kinetic energy in each calculation. For ions injected on-axis, the probability of ion capture is a flat-top function of the time when FEC is switched to close the trap. The width of this function is an estimate of the capture time width.  For the case of Ar\,{\scriptsize XIV}, CTW $\approx$ 80(20) ns is calculated for the optimal trapping conditions given above in Table \ref{tab: param}.  For comparison, in a high-field Penning trap with a long electrode stack, the ions are captured in a nearly-flat bottom (square-well) potential and the CTW is well approximated by the round-trip time, which can range from $\approx$ 300 ns \cite{rsiatrap} to about $1\, \mu$s \cite{retrap1}. The CTW of a compact Penning trap tends to be shorter due to its size.  However, as illustrated in this work, the CTW of a unitary Penning trap is sufficient to capture a broad range of ions.

\section{Experiments}
\label{sec: expres}

\subsection{Pulsed extraction of ions}
\label{sec: EBIT}
The energy available for electron impact ionization in an EBIT is set by a common-mode, float voltage applied to the drift tube assembly.  In this work, the float voltage is adjusted to give an electron beam energy ($E_{e-}$) in the range from 2.0 keV to 4.0 keV with an electron beam current ($I_{e-}$) in the range from 6 mA to 150 mA.  The NIST EBIT ion-extraction beamline has been optimized for high ion flux \cite{PikinBeamline} in ion-surface bombardment experiments \cite{hcibprl2011}, wherein the EBIT is typically operated in a continuous, high-current mode with $I_{e-} = 150$ mA.  For the ion capture experiments discussed here, it would be ideal for the extracted ions to be bunched tightly in both space and time.  Therefore, the EBIT is operated in a low-current, pulsed extraction mode.  The electron beam energy and current are chosen to optimize the production and capture of selected ions.  As an example we present the case of Ar$^{13+}$ extracted at an electron beam energy of E$_{e-} =$ 2.50 keV and electron beam current I$_{e-} =$ 14.4 mA.

To extract ions in pulses, a fast (rise time $\approx$ 50 ns) voltage pulse of 0 V to 400 V is applied to the MDT electrode in addition to the float voltage.  As indicated in Table \ref{tab: param}, the UDT electrode is biased at a lower potential than the LDT electrode.  Consequently, the rapid rise in MDT voltage pushes all ions in the EBIT into the beamline.  As illustrated in Figure \ref{fig: ebitapp}, ions leaving the EBIT are transported via the ion optics in the horizontal beamline to an analyzing magnet that filters to select a specific charge state. 

\begin{figure}[h]
\begin{center}
\includegraphics[angle=0,width=0.4\textwidth]{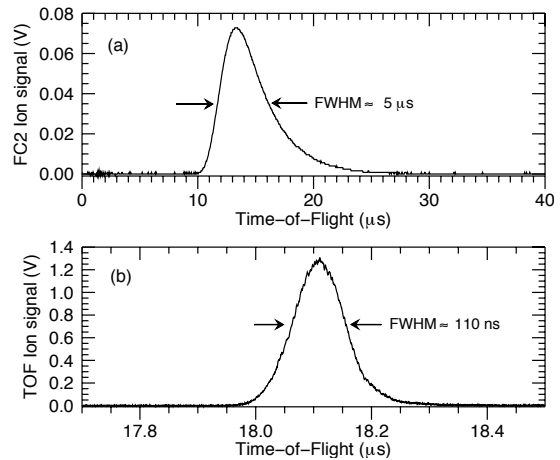}
\caption{\label{fig: EBIT-TOF} Detection of extracted ion bunch: (a) using a Faraday cup (FC2) before the analyzing magnet; and (b) using a fast TOF detector after selection of one charge state (Ar\,{\scriptsize XIV}) which is propagated through the Penning trap.  The detected Ar ions were produced with an electron beam energy (E$_{e-}$) and current (I$_{e-}$) of 2.50 keV and 14.4 mA, respectively.}
\end{center}
\end{figure}

Figure \ref{fig: EBIT-TOF} (a) shows a typical Faraday cup signal generated by ions of various charge states striking FC2 immediately in front of the analyzing magnet.  The analyzing magnetic field is tuned to single out a specific charge state to pass through the magnet, with its trajectories bent into the vertical beamline segment while all other charge states will hit the chamber wall.  Illustrative examples are provided in \cite{CDAMOP2011}.  The selected charge state is guided further into the ion capture apparatus.   For beam diagnostics, the extracted ion pulse passes through the grounded Penning trap and is detected using a fast TOF detector.  As shown in Fig \ref{fig: EBIT-TOF} (b), the charge-state-selected ion signal amplitude is $\approx$ 1.3 V and has a full width at half maximum (FWHM) of $\approx$ 110 ns, corresponding to $\approx$ 1435 ions per extraction pulse passing through the trap.  By fine tuning the electrostatic elements in the ion beamline, the EBIT settings, and the analyzing magnet field, this TOF signal is optimized for maximum ion pulse amplitude and minimum time width.

\subsection{Slowing and capture}
\label{sec: ioncap}

\begin{figure}
\begin{center}
\includegraphics[angle=90,width=0.4\textwidth]{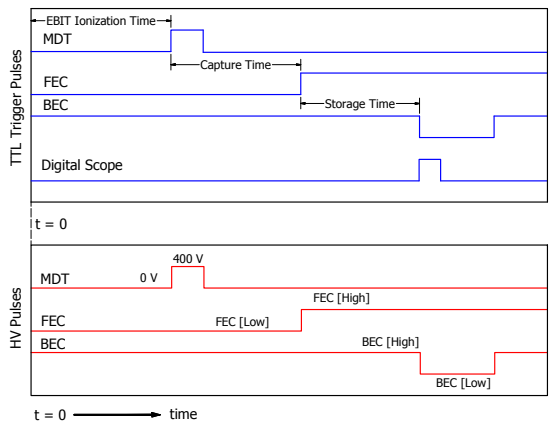}
\caption{\label{fig: puldiag}(Color online) Timing pulse diagram for controlling ion capture and detection.  TTL pulses triggering various switches/scopes are shown in the upper section (blue); corresponding high voltage outputs are shown in the lower section (red). Stored ions are ejected to a detector when BEC is low.  A schematic diagram for TOF detection is given in \cite{unitpen} and an abridged timing scheme is shown in \cite{CDAMOP2011}}
\end{center}
\end{figure}

Capturing the extracted ion pulse involves two key aspects: (1) closing the trap at the right time; and (2) tuning the float potential ($V_C$) of the unitary Penning trap to match the EBIT extraction energy.  The timing diagram for ion extraction and capture is shown in Fig. \ref{fig: puldiag}.  Details of the ion detection scheme are discussed in \cite{unitpen, CDAMOP2011}.

\begin{figure}
\begin{center}
\includegraphics[angle=0,width=0.5\textwidth]{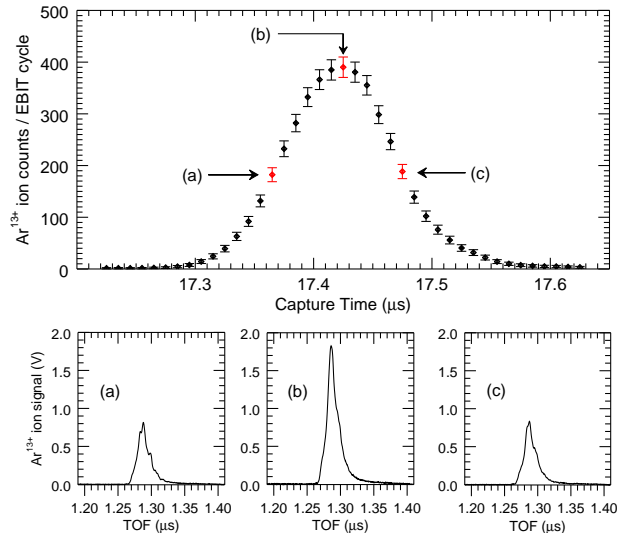}
\caption{\label{fig: captcomp}(Color online) Observed ion capture time profile for Ar\,{\scriptsize XIV}. Ion counts obtained by integrating TOF signals, as illustrated with 3 cases: (a) capture time below optimal value; (b) capture time at the optimal value; and (c) capture time above optimal value.  The TOF signals associated with these 3 cases (in red) are shown in three inset plots and labelled (a-c) correspondingly.  Ions were stored for 1 ms; the data represent the average of 64 trials each. Error bars represent 1 standard deviation.}
\end{center}
\end{figure}

Experimentally, the ``capture time,'' the time at which the entrance endcap electrode is switched to close the trap, is varied to maximize the number of ions captured per pulse.  A measurement of the optimal ion capture time is shown in Fig. \ref{fig: captcomp}.  Ions are captured and stored for 1 ms before being counted by ejection to the TOF detector.  In contrast to the ideal case presented in \S \ref{sec: ioncapsim}, the observed ion capture time profile (Fig. \ref{fig: captcomp} top) is mainly shaped by the characteristics of the ion pulse extracted from the EBIT. The observed peak gives the optimal capture time. In the case of Ar$^{13+}$ ions, the optimal capture time occurs at 17.43 $\mu$s after pulsed extraction from the EBIT with a nominal energy of 2.50 keV.

Another important consideration that affects the residual energy of captured ions is the deceleration of the ion pulse as it approaches the Penning trap, which is controlled largely by the common-mode, float voltage $V_C$ applied to all electrodes in the Penning trap assembly. In the continuous extraction mode, ions escape into the beamline with an energy of $E_{ion} \approx QU_{e-beam}$, where $U_{e-beam}$ is the electron beam energy; in contrast, for pulsed extraction mode, the fast switching of the MDT electrode gives ions an additional $\approx 400 \,Q$ eV of kinetic energy. The float voltage on the unitary Penning trap is adjusted to match the incoming ion energy, thus fine-tuning the amount of energy that is to be removed from the ion bunch in the process of being slowed and captured.  

\begin{figure}
\begin{center}
\includegraphics[angle=0,width=0.4\textwidth]{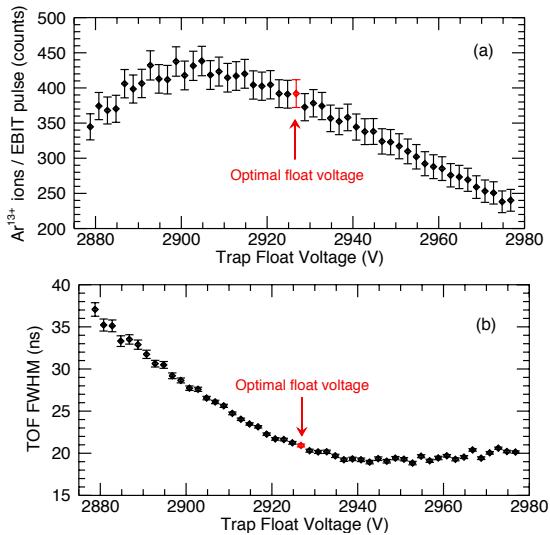}
\caption{\label{fig: trfloat}(Color online) Optimization of the common-mode, float voltage ($V_C$) applied on the compact Penning trap.  Figure (a) shows the number of ions detected as a function of float voltage, following 1 ms of ion storage, averaged over 64 pulses.  Figure (b) shows the TOF width of the ejected ion pulse.  The applied trap well is $V_o = 30$ V, and the capture time is $t_{capture} = 17.43 \mu$s. Error bars represent 1 standard deviation.}
\end{center}
\end{figure}

\begin{figure}
\begin{center}
\includegraphics[angle=0,width=0.4\textwidth]{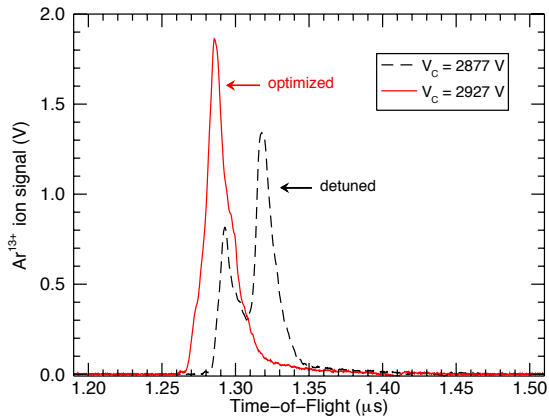}
\caption{\label{fig: tofwidth}(Color online) Optimized TOF signal from captured Ar\,{\scriptsize XIV}.  Captured ions are ejected after 1 ms of storage in the Penning trap. The narrow TOF signal in red solid line is for optimized float voltage $V_C$ = 2927 V, highlighted in Figure \ref{fig: trfloat}. For comparison, a double-peaked TOF signal corresponding to a detuned float voltage is also shown in black dashed line. Optimal capture time $\approx 17.43 \mu$s is used (see Figure \ref{fig: captcomp}). } 
\end{center}
\end{figure}

The influence of energy matching is illustrated in Figures \ref{fig: trfloat} and \ref{fig: tofwidth}.  The trap float voltage $V_C$ is adjusted to obtain the optimal ion capture signal.  The number of ions following 1 ms storage is measured as a function of the trap float voltage.  There is a broad maximum between 2880 V and 2940 V.  However, the width of the TOF signal drops steadily over that same voltage interval.  The narrowing of the TOF width as a function of the float voltage indicates that as $V_C$ is increased, the energy matching between the Penning trap and the extraction energy of the incoming ion pulse is improving.  As $V_C$ is further increased, the number of captured ions begins to decrease significantly, because more of the incoming ions lack the kinetic energy to reach the trapping region.

Dramatic broadening in the TOF signal for ions ejected from the Penning trap can result from mistuning of the float voltage, as illustrated in Figure \ref{fig: tofwidth}.
For a float voltage that is well below optimal value, the captured ions can have energy significantly higher than the bottom of the potential well, and a double peak structure in the TOF signal is observed.  For float voltages near the  optimal value, the TOF signal is single peaked and narrower, with an optimal FWHM $\approx$ 18.5 ns.  It is important for the TOF signal to be single peaked for proper interpretation of lower charge states generated after long storage times \cite{unitpen}.  Furthermore, as the float voltage approaches the optimal value from below, the TOF signal becomes narrower (see Figure \ref{fig: trfloat} b) indicating that the captured ions have less residual energy. 

\subsection{Energy of captured ions}
\label{sec: ionceng}

Experiments and model simulations, discussed in previous sections, have been useful in developing a unitary Penning trap for capturing multi-charged ions.  Trap parameters were deliberately sought to favor computed ion trajectories which lead to bound motions with small amplitudes. Furthermore, the control settings of the ion source, electrostatic ion optics, and compact Penning trap have been tuned in an attempt to maximize the number of ions captured, as well as to minimize the width of the time-of-flight signal.  Consequently, Fig. \ref{fig: trfloat}b indicates that the residual energy in bound ion motions can be significantly reduced.

To measure the energy distribution of captured ions, we used an over-the-barrier technique that is well-established in high-magnetic-field, multi-well Penning traps \cite{atrapiont}.  In the standard method, ions escaping from confinement are guided by strong magnetic field lines to an ion counter if they have sufficient energy to surmount a controlled potential barrier. The ion count is correlated with the instantaneous height of the potential barrier to obtain the energy distribution.   

\begin{figure}
\begin{center}
\includegraphics[angle=0,width=0.4\textwidth]{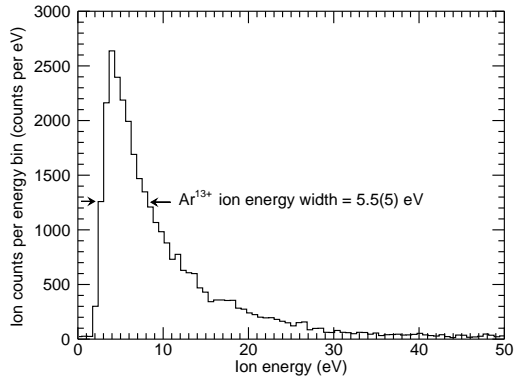}
\caption{\label{fig: iontemp} Observed energy distribution of Ar\,{\scriptsize XIV} ions escaping the confinement barrier along the trap axis as the ring electrode voltage is ramped linearly to shallower well depths. The energy width at half-maximum is 5.5(5) eV. Measured after 1 ms of storage.}
\end{center}
\end{figure}

The use of this method in a unitary Penning trap, on the other hand, requires some modification because of several features: (1) the magnetic field (maximum 0.31 T at the center) drops rapidly, particularly as the ions enter the endcap; (2) the reentrant endcaps make the well minimum very sensitive to asymmetrically applied voltages; (3) the ions are guided mainly by electrostatic ion optics to the detector.  Hence, in order to minimize the transport losses during the energy measurement, the ring electrode has been used to control the barrier height. The ion cloud energy, 1 ms after capture, has been measured by slowly ramping up the trap ring electrode voltage at a specified rate. As the ring voltage rises, the axial potential well depth decreases, allowing successively slower ions to escape over a known potential barrier in transit to the detector.  An ion energy distribution of Ar$^{13+}$ ions escaping from a unitary Penning trap is shown in Fig. \ref{fig: iontemp}.

The TOF detector was operated in the ion-counting mode, with a fully-saturated bias voltage of -1730 V.  A fast multichannel scaler was used to count events, triggered to begin acquisition simultaneously with the ramping of the ring electrode voltage. Since the ring electrode voltage is ramped at a controlled rate of $V_r (t) = $ 0.1V / $\mu$s $\times$ t, we can convert the arrival time of ions at the TOF detector to the corresponding ring electrode voltage, and hence to the barrier height.   An ion escaping along the trap axis must have energy exceeding $Q \, e \, \Delta V$ to surmount the barrier potential $\Delta V = \Delta V_0 - 0.388 V_r (t)$ where $\Delta V_0$ is the depth of the electrical potential well (maximum $-$ minimum) on axis.  For the case considered (Table \ref{tab: param}), $\Delta V_0 = 0.388 \times 30 V = 11.64 V$.

The energy distribution of Ar\,{\scriptsize XIV} ions escaping from the unitary Penning trap along its axis has a FWHM energy width of 5.5 $\pm$ 0.5 eV.  This energy distribution is a factor of $\approx$ 60 narrower than expected inside an EBIT \cite{Lapierre2005}. The over-the-barrier method generally gives an upper limit for the ion energy since the escaping ions tend to heat up from release of the ion cloud space-charge potential energy \cite{atrapiont}.  It is worth noting also that this is an estimate of the residual energy distribution shortly after capture, before any active cooling scheme has been implemented.  

Generally, a narrower energy distribution is favorable for spectroscopy because the Doppler broadening of spectral lines tend to have a Gaussian distribution with a FWHM line-width that is related to system parameters by $\Delta f_{\rm FWHM} = 2 f_o \sqrt{(2kT/Mc^2){\rm ln2}}$ where $f_o$ is the transition frequency, $k$ is the Boltzmann constant, $T$ is the ion cloud temperature, $M$ is the mass of the radiator, and $c$ is the speed of light.\cite{Griem1997} For example, the spectral lines emitted by an Ar$^{13+}$ ion cloud with temperature $kT \approx 5.5$ eV are expected to have a fractional Doppler line-width of $\Delta f/f_o \approx 2 \times 10^{-5}$.

\section{Ion capture efficiency}
\label{sec: capeff}

The number of extracted ions captured in the Penning trap is determined in part by the fixed parameters chosen for the trap and beam-tuning structures ({\it e.g.}, sizes of apertures); it is also affected by adjustments in operating conditions made during experiments to optimize energy and ion pulse width. Trade-offs are made in optimization, as illustrated in Fig.\ref{fig: trfloat}.  Assuming an incoming ion beam with no initial transverse momentum and neglecting space-charge effects, simulations show that ions arriving at a common time can be captured with 100 \% efficiency provided the beam radius is less than 2 mm.  In practice, the capture efficiency is observed to be roughly 60 \% largely because of the velocity spread in the extracted ion bunch.  Some ways of reducing the velocity spread to improve capture efficiency are described above.  In this section, we present measurements for estimating the number of stored ions and capture efficiency.

We measure the following quantities to characterize ion number in the Penning trap region: (a) $N_{FCA}$, the number of ions striking Faraday cup FCA after passing through the one-magnet Einzel lens with 11.11 mm inner diameter; (b) $N_{0V}$, the number of ions passing through the grounded trap and hitting the TOF detector; and (c) $N_{HV}$, the number of ions hitting the TOF detector after passing through the trap floated at high voltage $V_C$ but with the endcaps biased at low settings (Table \ref{tab: param}). Column 3 of Table \ref{tab: ioncapm} gives these measurements for extracted bunches of Ar$^{13+}$ ions. The number of ions determined from the Faraday cup signal $N_{FCA}$ is the largest since the ion bunch at FCA has not been partially clipped by the 8.00 mm diameter holes in the FEC and BEC electrodes. The active diameters of the FCA and TOF detectors are 9.525 mm and 8.00 mm, respectively.   

\begin{table}[htdp]
\squeezetable
\begin{center}
\begin{tabular}{|c|c|c|c|}
\hline
  & \multicolumn{3}{|c|}{Ar$^{13+}$ ion count} \\ \cline{2-4}
Detector (set-up) & symbol & Measured & Simulated \\
\hline
 & & & \\
FCA    (before trap) & $N_{FCA}$ & 5275 & 5275 \\
 & & & \\
TOF  (grounded trap) & $N_{0V}$ & 1435 & 1655 \\
 & & & \\
TOF (HV-biased trap) & $N_{HV}$ & 687 & 718 \\
 & & & \\
\hline
\end{tabular}
\end{center}
\caption[Ion pulse measurements]{Measurement of the number of Ar$^{13+}$ ions entering the trap region under three conditions.  $N_{FCA}$ is the number of ions measured on a Faraday cup before the trap.  $N_{0V}$ and $N_{HV}$ are the number of ions measured on the TOF detector when the Penning trap is fully grounded and floated for capture, respectively.}
\label{tab: ioncapm}
\end{table}

For comparison, we computed the ion transport for a Gaussian radial distribution of trajectories entering the one-magnet Einzel lens, passing through the trap, and terminating at the TOF detector. An initial ion velocity of $42\,840$ m/s is assigned entirely along the trap axis.   Previous experiment\cite{CDAMOP2011} has shown evidence to support a Gaussian density profile in a tightly-focussed beam.  The cross-sectional density is modeled by a Gaussian function:
\begin{equation}
\label{gausbm}
\sigma (r) = \frac{N_o}{2\pi R_B^2} \exp{\left({-\frac{r^2}{2R_B^2}}\right)}
\end{equation}
where $N_o$ is the total number of ions and $R_B$ is the one-sigma beam radius;  the number of ions within radius $r$ is given by the integral $N = \int_0^r 2 \pi r \sigma (r) \,dr$. The simulation results for $N_{o} = 5336$ ions and $R_B = 2.0$ mm are in the last column of Table \ref{tab: ioncapm}, and agree well with measurements (column 3) for the grounded trap and for the floated trap.  

For Ar$^{13+}$, Figures 9 and 10 indicate that about 400 ions were detected when the ion cloud in the Penning trap was ejected to the TOF detector.  To determine the capture efficiency for the Penning trap system, independent of the ion source and beamline used for production and transport of HCIs, we use the number of ions entering the Penning trap while at high voltage, $N_{HV}$, as the normalization.  The resulting efficiency is 57(16)\% for the Ar$^{13+}$ ion capture experiment.

This result agrees with a crude estimate of 61(10)\% for capture efficiency obtained from the simulations of Section \ref{sec: ioncapsim}.  Here the efficiency is calculated as the percentage of total ion signal that arrives at the TOF detector within $\pm$ CTW/2  of the TOF peak;  i.e.  $t_{peak} \pm 40$ ns in Fig. \ref{fig: EBIT-TOF}b.  For an on-axis beam, this is the maximum fraction of incoming ions that can be located inside the trap region at one time.

\section{Summary}
Highly-charged ions produced by electron impact ionization within an EBIT, with electron beam energy of a few keV, have been slowed and captured in a unitary Penning trap deployed on the existing ion-extraction beamline at NIST. The Penning trap is made very compact (less than 150 cm$^3$ in volume) by a unitary architecture that embeds two rare-earth permanent magnets within the electrode structure in order for the trapping apparatus to fit within space constraints, and to provide easy radial access to the stored ions. 

The procedure for capturing energetic ions in a unitary Penning trap is presented here with experimental results for the isolation of Ar$^{13+}$ ions, and is elucidated with simulations of single ion trajectories. Measurements confirm the importance of energy matching and precise timing of capture to achieve the lowest energy distribution for the isolated ions.  Simulations provide some insight in designing the set of conical, electrostatic decelerators near the entrance endcap of the ion trap to aid in maximizing ion capture and minimizing residual energy. As a demonstration, Ar$^{13+}$ ions extracted from the EBIT with $\approx$ 38 keV kinetic energy have been decelerated and captured with a residual energy spread of $\approx$ 5.5(5) eV, measured by ejecting the isolated ions to a TOF detector 1 ms after capture.  Without applying any active cooling, this observed energy distribution is $\approx 60$ times smaller than typically expected for ions inside an EBIT.  Colder ion clouds may be attainable by applying evaporative or sympathetic cooling techniques.  Recent theoretical studies propose various potential applications for isolated highly-charged ions, including optical frequency standards \cite{HCI-clock2012, Dzuba2012hci}, tests of fundamental symmetries \cite{Berengut2010prl}, and measurement of fundamental constants \cite{Mohr2008}.

\section{Acknowledgments}
Portions of this work were completed while Nicholas D. Guise held a National Research Council Associateship Award at NIST. We thank Yuri Ralchenko and Craig J. Sansonetti for reading this manuscript carefully and providing useful comments.


\begin{thebibliography}{40}
\bibitem [{dis()}]{disclaimer}%
  \BibitemOpen
  \href@noop {} {}\bibinfo {note} {Identification of a product herein is for
  documentation purposes only, and does not imply recommendation or endorsement
  by NIST, nor does it imply that this product is necessarily the best
  available for the purpose.}\BibitemShut {Stop}%
\end{thebibliography}
\end{document}